\documentclass[pra,aps,twocolumn,showpacs]{revtex4}
\usepackage{graphicx}
\usepackage{dcolumn}

\begin{document}
\def\be{\begin{equation}}
\def\ee{\end{equation}}
\def\bearr{\begin{eqnarray}}
\def\eearr{\end{eqnarray}}
\def\la{\langle}
\def\ra{\rangle}
\def\l{\left}
\def\r{\right}

\title{Collective modes of Fermi superfluid containing vortices along 
the BEC-BCS crossover}

\author{Tarun Kanti Ghosh and Kazushige Machida}
\affiliation
{Department of Physics, Okayama University, Okayama 700-8530, Japan}
\date{\today}

\begin{abstract}
Using the coarse-grain averaged hydrodynamic approach, we calculate
all low energy transverse excitation spectrum of a rotating Fermi 
superfluid containing vortex lattices for all regimes along the BEC-BCS 
crossover. 
In the fast rotating regime, the molecular BEC enters into
the lowest Landau level, but the superfluid in the unitarity and
the BCS regimes occupies many low-lying Landau levels.
The difference between the breathing mode frequencies at the
BEC and unitarity limit shrinks to zero as the rotation speed 
approaches the radial trap frequency,
in contrast to the finite difference in the non-rotating systems.
\end{abstract}

\maketitle

Strongly interacting two-component Fermi gases provide a unique
testing ground for the theories of exotic systems in nature.
In atomic Fermi gases, tunable strong interactions are produced
using Feshbach resonance \cite{houb,stwa,ties}.
By sweeping the magnetic field in the Feshbach resonance experiments,
magnitude and nature of the two-body interaction strength changes
from repulsive to attractive. Across the resonance the $s$-wave 
scattering length $a$ goes from large positive to large negative values.
The fermionic system becomes molecular Bose-Einstein condensates 
(BEC) for strong repulsive interaction and transform into the 
Bardeen-Cooper-Schrieffer (BCS) superfluid when the interaction is
attractive. The first observations of BEC of molecules
consisting of loosely bound fermionic atoms \cite{greiner,jochim,zw}
initiated a series of explorations \cite{hara,regal,bourdel1,barten,bourdel,chin}
of the crossover between BEC and BCS superfluid.
Near the resonance, the zero energy $s$-wave scattering length $a$
exceeds the interparticle spacing and the interparticle interactions
are unitarity limited and universal.

One of the hallmarks of superfluidity, be it bosonic or fermionic,
is the presence of quantized vortices.
Recently, large number of vortices in a Fermi gas along the BEC-BCS
crossover are indeed observed \cite{ketterle} and it proofs that this 
strongly interacting system is really superfluid. 
Equilibrium properties of vortices in a rotating Fermi superfluid 
described by BCS theory have already been the subject 
of several studies \cite{feder,feder1,bulgac1,tem,castin}.
There have been known several outstanding vortex problems
associated with the fermionic excitations around a vortex
core in the quantum limit of a superconductor under magnetic 
field \cite{haya}.

The successful measurements of the breathing mode frequencies in
a non-rotating Fermi superfluid \cite{kinast,bar} and 
the availability of the vortex lattices in a rotating Fermi
superfluid \cite{ketterle} have stimulated us to study theoretically 
the collective excitation spectrum of a rotating Fermi superfluid 
containing large number of vortices along the BEC-BCS crossover.
So far, both theoretical \cite{cozzini,cozzini1,bigelow} and 
experimental \cite{codd} results have been
obtained for low-energy modes of a rotating, weakly interacting 
condensates with large number of vortices. 
As in the case of a rotating bosonic clouds, the frequencies of the 
collective modes of a Fermi superfluid along the crossover in presence 
of the vortex lattices can be measured experimentally with high accuracy.

In this work, we calculate all low energy frequencies of the transverse  
excitations of a strongly interacting Fermi superfluid contains 
vortex lattices along the BEC-BCS crossover. 
In particular, we find that the Fermi superfluid in the BEC regime
enters into the lowest Landau level at fast rotation.
However, the superfluid in the unitarity and the BCS regimes 
occupy many low-lying Landau levels even at fast rotation. At 
fast rotation, the radial breathing
mode frequency becomes close to $ 2 \omega_r $ ($\omega_r $ is the
radial trap frequency) in all the regimes 
along the crossover, in contrast to the non-monotonous behavior of the breathing
mode along the crossover in the non-rotating systems \cite{stringari}.

We consider a large number of vortices formed in a rotating 
Fermi superfluid confined by a cigar shaped harmonic trap potential 
$ V_{\rm ho}(r,z) = (M/2)(\omega_r^2 r^2 + \omega_z^2 z^2) $.
We shall use the concept of diffused vorticity which has been
successfully used to study the low energy excitations and the dynamics
of the weakly interacting atomic BEC in presence of large number of vortices
\cite{cozzini,cozzini1,bigelow}. 
The diffused vorticity, $ {\bf \nabla } \times {\bf v} = 2 {\bf \Omega}$,
of a vortex lattice is an effective coarse-grain averaged description
that simplifies the formalism by smoothing out the effect of
the individual vortices. 
The concept of diffused vorticity is sufficient
to describe the dynamics at macroscopic distances, larger than the
average distance between vortices.

We assume that the system behaves hydrodynamically throughout all the regime.
If the system is BCS superfluid, then as long as  the oscillation frequency
is below the gap frequency ($\Delta_g/\hbar $) needed to break up a Cooper pair, 
this condition is expected to be fulfilled.
Within the coarse-grain hydrodynamic description, 
the equations of motion for the density and the velocity
field which are written by the following continuity and Euler 
equations, respectively,
\be
\frac{\partial n}{\partial t}  =  - {\bf \nabla} \cdot [n(r) {\bf v}],
\ee
and
\bearr
M \frac{\partial {\bf v} }{\partial t} & = &
- {\bf \nabla} [\frac{1}{2} M {\bf v}^2 + V_{\rm ho}(r,z) - 
\frac{1}{2} M \Omega^2 r^2 + \mu (n)]
\nonumber \\
& + & 2 M {\bf v} \times {\bf \Omega }
 +  M {\bf v} \times {\bf \nabla} \times {\bf v},
\eearr
where $ \mu(n) $ is the density-dependent chemical potential.
Also, $ {\bf v} = (\hbar/2M) \nabla \theta $ is the superfluid
velocity of a particle of mass $M$.
The terms $ 2 M {\bf v} \times {\bf \Omega } $ and $
\nabla [(1/2) M \Omega^2 r^2] $ are
the Coriolis and the centrifugal forces, respectively.
The equation of state enters through the density-dependent
chemical potential.
We assume the power-law form of the equation of state as
$ \mu(n) = C n^{\gamma} $.
At equilibrium, the density profile takes the form
$ n_0(r) = (\mu/C)^{1/\gamma}( 1- \tilde r^2)^{1/\gamma} $,
where $ \mu = (M/2)(\omega_r^2 - \Omega^2) R_{\perp}^2 $ is the chemical
potential, $R_{\perp} $ is the radial size and $ \tilde r = r/R_{\perp} $.
We are interested to study the radial excitations, therefore, we neglect
the weak harmonic confinement along the $z$-direction.

We linearize the density, velocity field and the
equation of state $\mu(n) $ around
their equilibrium values as $ n = n_0(r) + \delta n $,
$ {\bf v} = {\bf v}_0 + \delta {\bf v} $ and 
$ \mu(n) = \mu(n_0) + (\partial \mu/\partial n)|_{n =n_0} \delta n $.
In the co-rotating frame, the equilibrium velocity field $ {\bf v}_0 = 0 $.
Thus we obtain the equations
of motion for the density and the velocity fluctuations in the
rotating frame as
\be \label{density}
\frac{\partial \delta n}{\partial t}  =  
- {\bf \nabla }_r \cdot [n_{0}(r) \delta {\bf v}],
\ee
and
\be \label{velocity}
M \frac{ \partial \delta {\bf v}}{\partial t}  =  
- {\nabla}_r[ \frac{\partial \mu(n) }{\partial n}|_{n =n_0}  \delta n ] +
2 M  \delta {\bf v} \times {\bf \Omega }.
\ee
We shall follow Ref. \cite{bigelow} to determine the relationship between 
the density and the phase fluctuations in presence of the vortex lattices.
We are looking for the normal mode solutions 
of the density and phase fluctuations as
$ \delta n \propto  e^{i m \phi} e^{-i\omega t}  $ and 
$ \delta \theta \propto  e^{i m \phi} e^{-i\omega t} $,
where $m$ is the azimuthal quantum number, $ \phi $ is the 
polar angle and $\omega $ is the excitation frequency in the
rotating frame.
Equation (\ref{velocity}) can be written in terms of the phase 
fluctuations after separating into radial and angular 
components as
\be \label{phase}
i \hbar \omega \frac{\partial \delta \theta } {\partial r}
= 2 \frac{\partial } {\partial r} [\frac{\partial \mu(n) }
{\partial n}|_{n =n_0} \delta n ] - 2 i \hbar \Omega \frac{m}{r} 
\delta \theta,
\ee
and
\be \label{phase1}
i \hbar \omega \delta \theta  = 2 \frac{\partial \mu(n) }
{\partial n}|_{n =n_0} \delta n - 2 i \hbar \Omega \frac{r}{m}  
\frac{\partial \delta \theta }{\partial r}.
\ee 
Integrating Eq. (\ref{phase}) with respect to $r$, we then get,
\be \label{equal}
i \hbar \omega  \delta \theta = 2 \frac{\partial \mu(n) } 
{\partial n}|_{n =n_0}\delta n - 2i m \hbar \Omega \int dr \frac{1}{r} 
\delta \theta.
\ee
Equating equations (\ref{phase1}) and (\ref{equal}), one gets
\be
\int dr \frac{1}{r} \delta \theta  - 
\frac{r}{m^2} \frac{\partial \delta \theta }{\partial r} = 0.
\ee
Differentiating the above equation on both sides with respect 
to the radial coordinate $ r $ and it gives the Cauchy equation:
\be
(r^2 \frac{\partial^2}{\partial r^2} 
+ r\frac{\partial}{\partial r} -m^2) \delta \theta = 0.
\ee
The general solution of the above equation is given by
$ \delta \theta = c^+ r^m  + c^- r^{-m}$,
where $ c^+ $ and $ c^- $ are arbitrary constants.
The velocity fluctuation along the radial direction can be obtained as
$ \delta v_r
= (\hbar/2M) (m/r) (c^+ r^m - c^- r^{-m})$.
The radial velocity can not diverge as $ r \rightarrow 0$. 
Therefore,
$ c^- = 0 $ for $m > 0$ and $ c^+ = 0 $ for $m < 0$.
This gives us
\be \label{solu}
\frac{\partial \delta \theta }{\partial r} = \pm \frac{m}{r} \delta \theta.
\ee
Substituting Eq. (\ref{solu}) into Eq. (\ref{phase1}), we obtain the 
following relationship between the density and the phase fluctuations:
\be\label{relation}
i( \hbar \omega \pm 2 \hbar \Omega) \delta \theta =  
2 \frac{\partial \mu(n) } {\partial n}|_{n =n_0} \delta n.
\ee

Taking gradient of Eq. (\ref{relation}) and then substituting 
it into Eq. (\ref{density}), we then get,
\bearr \label{main}
0 & = & \frac{\gamma}{2}(1-\tilde \Omega^2)
 \nabla_{\tilde r} \cdot [(1- \tilde r^2)^{1/\gamma}
\nabla_{\tilde r}(1- \tilde r^2)^{1-1/\gamma} \delta n(r)] \nonumber \\
& - & \tilde \omega_\alpha (\tilde \omega_\alpha \pm 2 \tilde \Omega) \delta n (r),
\eearr
where $ \tilde \omega = \omega/\omega_r $ and 
$ \tilde \Omega = \Omega/\omega_r $. Here, $\alpha: (n_r,m)$ is a set of 
two quantum numbers, where $n_r$ is the  radial quantum number.

When $ \Omega =0 $, the solutions of Eq. (\ref{main}) are 
known analytically \cite{tkg}. For $ \Omega = 0 $, the 
eigenspectrum is given by
\be
\tilde \omega_\alpha^2 = |m| + 2 n_r [\gamma(n_r + |m|) + 1],
\ee
and the corresponding eigenfunction is given by
\be
\delta n_\alpha (\tilde r,\phi) \propto (1-\tilde r^2)^{1/\gamma -1} 
\tilde r^{|m|} P_{n_r}^{(1/\gamma -1, |m|)}(2\tilde r^2 -1) e^{im\phi},
\ee
where $ P_{n}^{(a,b)}(x) $ is a Jacobi polynomial of order $n$. 
Assuming the net effect of a coarse grain averaged vortex lattice 
on the wave functions to be a modification of the effective trapping 
potential. Therefore, we shall use the above wave function as our ansatz.
With this ansatz for $\delta n(\tilde r,\phi) $, the expectation value of 
Eq. (\ref{main}) is given by 
\be
\tilde \omega_\alpha (\tilde \omega_\alpha \pm 2 \tilde \Omega) = 
[|m| + 2n_r \gamma (n_r + |m| + 1/\gamma )](1-\tilde \Omega^2).
\ee
The above quadratic equation can be solved to get the following 
eigenfrequencies in the rotating frame:
\be \label{freq}
\tilde \omega_\alpha = \sqrt{[|m| + 2n_r \gamma(n_r + |m| + 1/\gamma )]
(1- \tilde \Omega^2) + \tilde \Omega^2 } \mp \tilde \Omega.
\ee 
The eigen frequencies in the laboratory frame are obtained by 
using the simple relation $ \omega_{\rm rot} = \omega_{\rm lab} - m 
\Omega $ and it is given by
\bearr \label{freq1}
\tilde \omega_\alpha & = & \sqrt{[|m| + 2n_r \gamma(n_r + |m| + 1/\gamma )]
(1-\tilde \Omega^2) + \tilde \Omega^2 } \nonumber \\ 
& \pm & (|m|-1) \tilde \Omega.
\eearr
The frequencies of the $m=0 $ modes are identical in both the rotating and
the laboratory frames and the energy degeneracy is restored for the
$ m = \pm 1 $ modes in the laboratory frame. The dipole mode
gives $ \tilde \omega_{0,\pm 1 } = 1 $ for all $ \Omega $ 
which satisfies the generalized Kohn's theorem.

When $ \gamma = 1 $, it reproduces the known result for weakly
interacting rotating BEC with a large number of vortices 
\cite{bigelow}. The eigenfrequencies given in Eq. (\ref{freq1}) 
with $ \gamma = 1 $ successfully describes the low energy 
transverse excitations of a rotating atomic BEC containing large 
number of vortices \cite{bigelow}.
In particular, the first ($\omega_{1,0}$) and second 
($\omega_{2,0}$) radial breathing modes are in good agreement 
with the experimental as well as numerical results of weakly 
interacting BEC \cite{codd,bigelow1}.
Moreover, the quadrupole modes ($\omega_{0,\pm2}$) are in good 
agreement with the numerical results obtained by solving the 
Bogoliubov-de Gennes equation \cite{mizu}.

At zero temperature, the energy per particle of a dilute Fermi 
system can be written as
$ \epsilon = (3/5) E_F \epsilon (y) $,
where $ E_F = \hbar^2 k_F^2/2M $ is the free particle Fermi energy and
$ \epsilon (y) $ is a function of the interaction parameter $y = 1/k_F a $.
The Monte Carlo calculations of
Refs. \cite{carl,astra} find $ \epsilon (y \rightarrow 0) = 0.42 \pm 0.01 $
which is in fair agreement with the experiments \cite{bourdel1,barten}.
On the basis of the data of Ref. \cite{astra}, Manini and Salasnich 
\cite{manini} proposed the following analytical fitting
formula of $ \epsilon(y)$ for all the regimes along the BEC-BCS crossover: 
\be \label{fit}
\epsilon (y) = \alpha_1 - \alpha_2
\tan^{-1}[\alpha_3 y \frac{\beta_1 + |y|}{\beta_2 + |y|}].
\ee
This analytical expression is well fitted with the data of Ref. \cite{astra}
for a wide range of $y$ on both sides of the resonance. 
Two-different sets of parameters are considered in Ref. \cite{manini}: one
set in the BCS regime ($y<0$) and another set in the
BEC regime ($y>0$). In the BCS limit, the values of the parameters 
\cite{manini} are $ \alpha_1 = 0.42 $, $ \alpha_2 = 0.3692 $, 
$\alpha_3 = 1.044 $, $\beta_1 = 1.4328 $ and $ \beta_2 = 0.5523 $. 
In the BEC limit,
the values of the parameters \cite{manini} are
$ \alpha_1 = 0.42 $, $ \alpha_2 = 0.2674 $, $\alpha_3 = 5.04 $,
$\beta_1 = 0.1126 $ and $ \beta_2 = 0.4552 $.

The chemical potential $\mu $ is given by \cite{manini}
\be \label{chemical}
\mu = \epsilon(n) + n \frac{d\epsilon (n)}{d n} =
E_F [\epsilon (y) - \frac{y}{5} \epsilon^{\prime}(y)] = E_F F(y),
\ee
where $ \epsilon^{\prime}(y) = \partial \epsilon (y)/\partial y $
and $ F(y) = \epsilon (y) - (y/5) \epsilon^{\prime}(y) $.
Due to the effect of the solid body rotation and the harmonic 
trapping potential, the free Fermi energy can be written as 
$ E_F = \hbar \omega_r [3N \lambda (1- \tilde \Omega^2)]^{1/3} $
with $ \lambda = \omega_z/\omega_r $.
The radial and the axial sizes can be written, respectively, as
$ R_{\perp} = a_r (24 N \lambda)^{1/6} (1-\tilde \Omega^2)^{-1/3} \sqrt{F(y)}$ and
$ R_z = a_z (24 N/ \lambda^2)^{1/6} (1-\tilde \Omega^2)^{1/6} \sqrt{F(y)}$, where
$ a_i = \sqrt{\hbar/M\omega_i} $ is the harmonic oscillator length with
$i = r,z$. The aspect ratio is given by
$ R_z/R_{\perp} = (1-\tilde \Omega^2)^{1/2}/ \lambda $ which shows that the
rapidly rotating Fermi superfluid takes a pancake form even if the
confining trap has a cigar shape. It is interesting to note that
the aspect ratio is the same for all the regimes along the
BEC-BCS crossover. 
  
\begin{figure}[ht]
\includegraphics[width=9.1cm]{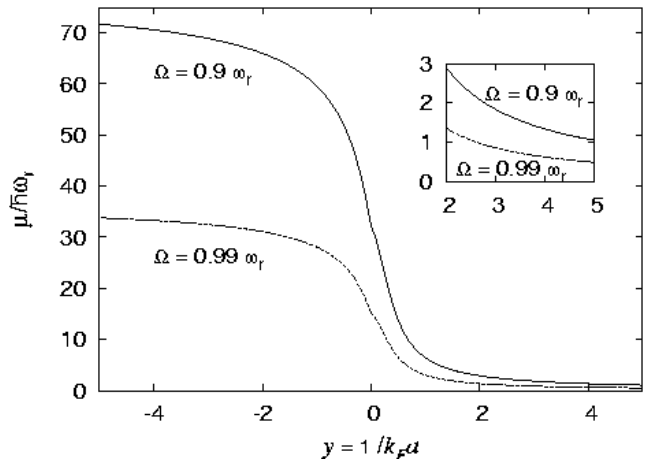}
\caption{Plots of the chemical potential vs coupling parameter $y$ 
of a trapped Fermi gas with $ N = 2 \times 10^6 $ and $ \lambda = 23/57 $.}
\end{figure}
It is seen from Eq. (\ref{freq1}) that when $ \Omega \rightarrow \omega_r$
all transverse modes with $ m =0$ in all the regimes along the
BEC-BCS crossover approach to the degenerate value $ 2 \omega_r $.
Such a macroscopic degeneracy is reminiscent of the
degeneracy of the energy levels of a single particle in an
uniform magnetic field, leading to the well known Landau level
structure. However, we should mention that our results here hold
not for a single particle, but for a system of interacting particles
along the BEC-BCS crossover. Therefore, the strongly interacting, rotating
Fermi superfluid forms the Landau level structure when
$ \Omega \rightarrow \omega_r $. 
In Fig. 1, we plot the chemical potential for various values of $ \Omega $.
Fig. 1 shows that the chemical potential
in the BEC side is less than the cyclotron energy gap ($ 2 \hbar \omega_r $)
at the fast rotation and the system enters into the lowest Landau level.
However, at the fast rotation, the chemical potential in the unitarity
and in the BCS regimes are quite large compared to the cyclotron energy gap.
The fermionic superfluid in the BCS as well as in the
unitarity regimes occupies many low-lying Landau levels even at
$ \Omega \rightarrow \omega_r $.

One can extract an effective adiabatic index $\gamma $ and its dependence
on the scattering length $a$ by defining the logarithmic derivative as 
\cite{manini}
\be \label{gamma}
\gamma \equiv \bar \gamma = \frac{n}{\mu}\frac{d\mu}{dn}
= \frac{\frac{2}{3} - \frac{2y}{5} \epsilon^{\prime}(y) +
\frac{y^2}{15} \epsilon^{\prime \prime}(y)}{\epsilon (y) -
\frac{y}{5}\epsilon^{\prime}(y)}.
\ee
\begin{figure}[ht]
\includegraphics[width=9.1cm]{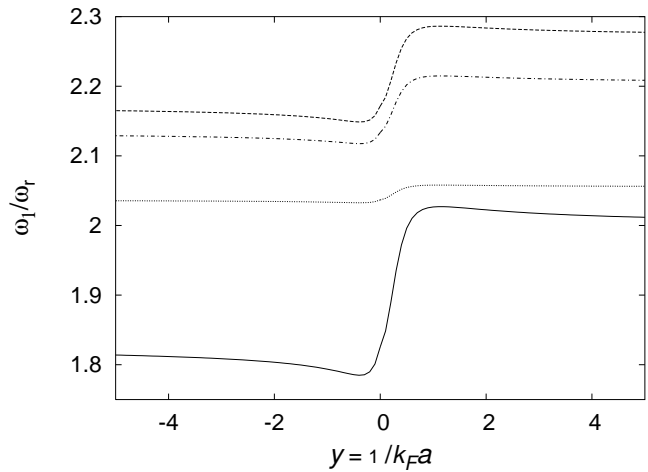}
\caption{Plots of the first breathing mode frequency for
$ \tilde \Omega = 0 $ (solid line), $ \tilde \Omega = 0.75 $ (dashed line),
$ \tilde \Omega = 0.85 $ (dot-dashed line) and $ \tilde \Omega = 0.97 $ (dotted line).}
\end{figure}

From Eq. (\ref{freq1}), the first radial breathing mode frequency can be written
as 
\be
\tilde \omega_1 = \sqrt{2(\gamma +1)(1- \tilde \Omega^2) + \tilde \Omega^2} 
+ \tilde \Omega.
\ee
When $ \Omega = 0 $, the difference between the breathing mode 
frequencies at the BEC limit and unitarity limit is 
$ \Delta(\Omega = 0 ) = \omega_1(\gamma = 1) - \omega_1(\gamma = 2/3) 
= (2 - \sqrt{10/3}) \omega_r = 0.174258 \omega_r$.
Recently, the radial breathing mode frequency of a non-rotating Fermi superfluid
along the crossover has been measured \cite{kinast,bar} which is in fair agreement
with the theoretical result $ \omega_1(\Omega = 0) = \sqrt{2(\gamma + 1)} \omega_r $.  
In Fig. 2, we plot the first radial breathing mode in all the regimes
along the BEC-BCS crossover for various values of $ \Omega $.
It is interesting to see from Fig. 2 that the difference between the 
breathing mode frequencies at the BEC-limit and the unitarity limit 
decreases as $ \Omega $ increases. In fact, the breathing mode in all the regimes
approaches $ 2 \omega_r $ when $ \Omega \rightarrow \omega_r $.
The shrinking of the difference of the breathing mode frequency could be
observed in the future experiment.

In conclusion, we have obtained an analytic expression for the collective excitation
spectrum of a Fermi superfluid containing vortices along the BEC-BCS crossover. 
The spectrum reduces to the Landau level structure in the fast rotating regime.
The superfluid in the BEC regime goes into the lowest Landau level when 
$ \Omega \rightarrow \omega_r $. However, the superfluid in the unitarity
and the BCS regimes occupies many low-lying Landau levels at the fast rotating regime.
The difference between the breathing mode frequencies at the unitarity and the 
BEC limits shrinks to zero in the fast rotating regime.

\begin{acknowledgments}
This work of TKG was supported by a grant (Grant No. P04311) of the
Japan Society for the Promotion of Science.
\end{acknowledgments}


\begin{thebibliography}{2}

\bibitem{houb}
M. Houbiers, H. T. C. Stoof, W. I. McAlexander, and R. G. Hulet,
Phys. Rev. A {\bf 57}, 1497(R) (1998).

\bibitem{stwa}
W. C. Stwalley, Phys. Rev. Lett. {\bf 37}, 1628 (1976).

\bibitem{ties}
E. Tiesinga, B. J. Verhaar, and H. T. C. Stoof, Phys. Rev. A {\bf 47}, 4114 (1993).

\bibitem{greiner}
M. Greiner, C. A. Regal, and D. S. Jin, Nature {\bf 426}, 537 (2003).

\bibitem{jochim}
S. Jochim, M. Bartenstein, A. Altmeyer, G. Hendl, S. Riedl, C. Chin, J. Hecker 
Denschlag, and R. Grimm, Science {\bf 302}, 2101 (2003).

\bibitem{zw}
M. W. Zwierlein, C. A. Stan, C. H. Schunck, S. M. F. Raupach, S. Gupta,
Z. Hadzibabic, and W. Ketterle, Phys. Rev. Lett. {\bf 91}, 250401 (1993).

\bibitem{hara}
K. M. O'Hara, S. L. Hemmer, M. E. Gehm, S. R. Grande,
and J. E. Thomas, Science {\bf 298}, 217 (2002).

\bibitem{regal}
C. A. Regal, M. Greiner, and D. S. Jin, Phys. Rev. Lett. {\bf 92}, 040403 (2004).

\bibitem{bourdel1}
T. Bourdel, J. Cubizolles, L. Khaykovich, K. M. F. Magalha£e, S.  J. J. M. F. Kokkelmans, 
G. V. Shlyapnikov, and C. Salomon, Phys. Rev. Lett. {\bf 91}, 020402 (2003).

\bibitem{barten}
M. Bartenstein, A. Altmeyer, S. Riedl, S. Jochim, C. Chin, J. H. Denschlag, and R.
Grimm, Phys. Rev. Lett. {\bf 92}, 120401 (2004).

\bibitem{bourdel}
T. Bourdel, L. Khaykovich, J. Cubizolles, J. Zhang, F. Chevy, M. Teichmann, L. Tarruell, S. J.
J. M. F. Kokkelmans, and C. Salomon, Phys. Rev. Lett. {\bf 93}, 050401 (2004).

\bibitem{chin}
C. Chin, M. Bartenstein, A. Altmeyer, S. Riedl, S. Jochim, J. H. Denschlag,
and R. Grimm, Science {\bf 305}, 1128 (2004).

\bibitem{ketterle}
M. W. Zweirlein, J. R. Abo-Shaeer, A. Schirotzek, C. H. Schunck, and W. Ketterle,
Nature {\bf 435}, 1047 (2005).

\bibitem{feder}
D. L. Feder, Phys. Rev. Lett. {\bf 93}, 200406 (2004).

\bibitem{feder1}
N. Nygaard, G. M. Bruun, C. W. Clark, and D. L. Feder, 
Phys. Rev. Lett. {\bf 90}, 210402 (2003).

\bibitem{bulgac1}
A. Bulgac and Y. Yu, Phys. Rev. Lett. {\bf 91}, 190404 (2003).

\bibitem{tem}
J. Tempere, M. Wouters, and J. T. Devreese, Phys. Rev. A {\bf 71}, 
033631 (2005).

\bibitem{castin}
G. Tonini and Y. Castin, cond-mat/0504612.

\bibitem{haya}
N. Hayashi, T. Isoshima, M. Ichioka, and K. Machida,
Phys. Rev. Lett. {\bf 80}, 2921 (1998).


\bibitem{kinast}
J. Kinast, S. L. Hemmer, M. E. Gehm, A. Turlapov, and J. E. Thomas,
Phys. Rev. Lett. {\bf 92}, 150402 (2004).

\bibitem{bar}
M. Bartenstein, A. Altmeyer, S. Riedl, S. Jochim, C. Chin, J. H. Denschlag,
and R. Grimm, Phys. Rev. Lett. {\bf 92}, 203201 (2004).



\bibitem{cozzini}
M. Cozzini and S. Stringari, Phys. Rev. A {\bf 67}, 041602(R) (2003).

\bibitem{cozzini1}
F. Chevy and S. Stringari, Phys. Rev. A {\bf 68}, 053601 (2003).

\bibitem{bigelow}
S. Choi, L. O. Baksmaty, S. J. Woo, and N. P. Bigelow,
Phys. Rev. A {\bf 68}, 031605(R) (2003).

\bibitem{codd}
I. Coddington, P. Engels, V. Schweikhard, and E. A. Cornell, 
Phys. Rev. Lett. {\bf 91}, 100402 (2003).

\bibitem{stringari}
S. Stringari, Europhys. Lett. {\bf 65}, 749 (2004).


\bibitem{tkg}
T. K. Ghosh and K. Machida, to appear in Phys. Rev. A.
Preprint: cond-mat/0510160.

\bibitem{bigelow1}
L. O. Baskmaty, S. J. Woo, S. Choi, and N. P. Bigelow,
Phys. Rev. Lett. {\bf 92}, 160405 (2004).

\bibitem{mizu}
T. Mizushima, Ph. D. Thesis, Okayama University,
Okayama, Japan.

\bibitem{carl}
J. Carlson, S.-Y. Chang, V. R. Pandharipande, and K. E. Schmidt,
Phys. Rev. Lett. {\bf 91}, 050401 (2003).

\bibitem{astra}
G. E. Astrakharchik, J. Boronat, J. Casulleras, and S. Giorgini,
Phys. Rev. Lett. {\bf 93}, 200404 (2004).

\bibitem{manini}
N. Manini and L. Salasnich, Phys. Rev. A {\bf 71}, 033625 (2005).

\end{thebibliography}
\end{document}